\documentclass[useAMS,usenatbib]{mnras}
\usepackage{amsmath}
\usepackage{graphicx,txfonts,fleqn,enumerate,color}

\renewcommand*{\vec}[1]  {\boldsymbol{#1}}
\renewcommand*{\v}[1]  {\boldsymbol{#1}}

\newcommand*  {\Exp}[1]  {\mathrm{e}^{#1}}

\newcommand*  {\Vp}      {V_{\mathrm{pl}}}

\newcommand   {\sub}[2]  {{#1}_{\mathrm{#2}}}

\newcommand*  {\twovector}[2] {{\begin{pmatrix} $1 \\ $2 \end{pmatrix}}}

\renewcommand {\emph}[1]  {\textit{#1}}

\title[Improvement in simulated orbits from smoothness]
{{Improving the accuracy of simulated chaotic $N$-body orbits using smoothness}}

\author[David M. Hernandez]
	{David M. Hernandez$^{1,2}$ \thanks{{{Email: dmhernandez@cfa.harvard.edu} }}  \\
	$^1$ Harvard--Smithsonian Center for Astrophysics, 60 Garden St., MS 51, Cambridge, MA 02138, USA \\
	$^2${RIKEN Center for Computational Science}, 7-1-26 Minatojima-minami-machi, Chuo-ku, Kobe, 650-0047 Hyogo, Japan \\
	}
\begin{document}

\maketitle

\label{first page}
\begin{abstract} 
Symplectic integrators are a foundation to the study of dynamical $N$-body phenomena, at scales ranging from from planetary to cosmological.  These integrators preserve the Poincar\'{e} invariants of Hamiltonian dynamics.  The $N$-body Hamiltonian has another, perhaps overlooked, symmetry: {it is smooth, or, in other words, it has infinite differentiability class order (DCO)} for particle separations greater than $0$.  {P}opular symplectic integrators, such as hybrid methods or block adaptive stepping methods {do not come from smooth Hamiltonians} and it is perhaps unclear whether they should.  We investigate the importance of this symmetry by considering hybrid integrators, whose {DCO} can be tuned easily.  Hybrid methods are smooth, except at a finite number of phase space points.  We study chaotic planetary orbits in a test considered by Wisdom.  We find that increasing {smoothness}, at negligible extra computational cost in particular tests, improves the Jacobi constant error of the orbits by about $5$ orders of magnitude in long-term simulations.  The results from this work suggest that smoothness of the $N$-body {Hamiltonian} is a {property} worth preserving in simulations.
\end{abstract}
\section{Introduction}
\label{sec:intro}
Symplectic integrators have been successful at studying a wide range of $N$-body particle dynamics in astrophysics, from planetary systems to galaxies with dark matter.  They are usually severely limited in that they require a timestep, but gravitation has no timescales associated with it.  To combat this limitation, a number of techniques have been employed \citep{H19} such as block timestepping or hybrid integration, which have become the basis of popular dynamics codes.  These modifications change the smoothness of the {$N$-body integrator Hamiltonian}.  The $N$-body Hamiltonian is a $C^\infty$ function for the domain of separations greater than $0$, meaning it can get differentiated infinite times with respect to phase space coordinates and still be {continuous}.   {It is also called smooth.  Smoothness measures the number of continuous derivatives.}  Traditional symplectic integrators constructed through operator splitting \citep{Y90,WH91} respect this symmetry of the $N$-body Hamiltonian and are {obtained from} $C^\infty$ {Hamiltonians}.  However, hybrid integrators \citep{C99,Reinetal2019,H19,Wisdom2017} are {derived from $C^n$ Hamiltonians}, where $n$ is an integer satisfying $n \ge 0$.  A $C^n$ function can be differentiated $n$ times and still be continuous.  {$n$ is denoted the differentiability class order (DCO).}  {A $C^{1,1}$ function has a Lipschitz continuous, rather than continuous, derivative.}  Block time-stepping methods \cite{farr07} and integrator subcycling methods \cite{S05} are not derived from smooth Hamiltonians; they switch in analogy to the `modified Heaviside function' from \cite{Reinetal2019}.  It remains to be seen whether Hamiltonian smoothness is a symmetry which integrators, whether they are symplectic or not, should mimic, just as they might mimic the conservation of Poincar\'{e} invariants,  the existence of a global Hamiltonian \citep[Section IX.3.2]{hair06}, the time-reversibility \citep{HB18}, or the angular and total momentum conservation.  {If an integrator is not Hamiltonian, we can still quantify its smoothness through its modified differential equations \citep[e.g.][]{hair06,HB18}}.  

\cite{H19} has explored the question of integrator smoothness for regular orbits.  The work considered hybrid integrators, because it is simple to modify {the smoothness of the Hamiltonian from which they are derived, which we define as the smoothness of the integrator.}  Hybrid integrators are $C^\infty$ except at a finite number of position coordinates.  Consistent with the predictions of \cite{AX16}, \cite{H19} found that{, apart from caveats,} smoothness of at least $C^{1,1}$ yielded periodic orbits.  Increasing the smoothness past $C^{1,1}$ did not {help bound the error}.

The situation is different for chaotic orbits.  \cite{Reinetal2019} studied the effect of Hamiltonian smoothness during a short-term chaotic simulation which integrates over discontinuities twice or four times, except in the modified-Heaviside case.  The errors were studied using approximations to commutators which may or may not converge.  \cite{Reinetal2019} argued that over this short-term simulation, Hamiltonian smoothness made no significant impact.  However, the advantage of symplectic integrators is in their long-term behavior.  

In this work, we study whether the {smoothness of the $N$-body Hamiltonian matters for simulating long-term chaotic orbits.}  We consider hybrid integrators, as in \cite{H19}, because their smoothness is easy to tune.  We construct a hybrid $N$-body integrator combining ideas from \cite{C99}, \cite{Reinetal2019}, and \cite{Wisdom2017}.  According to the modified differential equations (MDEs) {\citep[e.g.][]{HB18,hair06}}, we predict Hamiltonian smoothness should have a significant impact on the error of integrators.  To test these ideas, we consider a restricted three-body problem in which the test particle executes a chaotic orbit in which it can visit both masses.  We observe integrator performance improvement by increasing the {DCO} from $n = 0$ to $n = 4$.  The error of the integrator apparently saturates for higher $n$.  The MDEs suggest that the effects of nonsmoothness, for this particular hybrid integrator, should disappear as the timestep approaches 0.  We verify this numerically by rerunning our test with a step $10$ times smaller.  {We also find that increasing the DCO improves the error in regular orbits.}  This work provides evidence that the smoothness of the $N$-body problem is a {feature} which should be followed as closely as possible by $N$-body integrators.  This is not the case for the other methods described above such as block time-stepping.

In Section \ref{sec:algo} we show how we construct the hybrid $N$-body integrators.  We carry out error analysis of a similar hybrid integrator using MDEs in Section \ref{sec:MDE} and give an example of how a non-smooth function leads to secular error.  Numerical experiments demonstrating the significance of Hamiltonian smoothness are carried out in Section \ref{sec:numex}.  Section \ref{sec:conc} concludes and discusses the impact of this work on $N$-body {integrations.}
\section{Algorithm description}
\label{sec:algo}
To construct a hybrid $N$-body integrator, we combine ideas from \cite{C99} and \cite{Wisdom2017}.  A coordinate system must be chosen.  {As pointed out by \cite{C99}, Jacobi coordinates are problematic if the planets have mass.}  Jacobi coordinates are appropriate if massive bodies cannot have close encounters, but massless test particles can have close encounters with the massive bodies.  This was the situation studied by \cite{Wisdom2017}.  We thus use Democratic Heliocentric coordinates {\citep[e.g.][]{HD17}}, which accommodate both massive and massless possibly having close encounters.  The Hamiltonian in these coordinates is split into two parts, $A$ and $B$, given by \citep{C99}:
\begin{subequations} \label{eqs:split1}
\begin{eqnarray}
	{A}{} &= &  
		T_1 + V_{\sun} + {\sub{V}{p1}} \qquad\text{and}
	\\ \label{eq:WHD:BB}
	{B}{} &=&
		T_0 +  {\sub{V}{p2}}.
\end{eqnarray}
\label{eq:Hsplit}
\end{subequations}
We defined,
\begin{equation} 
\begin{aligned}
\label{eq:defterms}
	T_1 &= \sum_{i\neq0} \frac{\vec{P}_i^2}{2m_i},
	\qquad
	V_\odot = -\sum_{i\neq0}\frac{Gm_0m_i}{Q_i}, \\&
	{\sub{V}{p1}} = - \sum_{0<i<j} \frac{Gm_im_{\!j}}{Q_{i\!j}} \left(1 - K(Q_{i j})\right),  \\&
	T_0 = \frac{1}{2m_0}\left(\sum_{i\neq0}\vec{P}_i\right)^2,
	\qquad\text{and}\qquad
	{\sub{V}{p2}} = - \sum_{0<i<j} \frac{Gm_im_{\!j}}{Q_{i\!j}} K(Q_{i j}).
	\end{aligned}
\end{equation}

$K$ is a function such that $0 \le K \le 1$.  $A$ is Liouville integrable if all $K$ equal $1$.  Then, a Kepler advancer such as that of \cite{WH15} can be used to solve it.  Otherwise, $A$, is not described by periodic orbits, and numerical approximation methods, such as those relying on computing Taylor series solutions to the ordinary differential equations (ODEs) to high powers in an expansion parameter must be used{.  Such a method is that of \cite{RS15}, which uses substeps to calculate series coefficients.}  $B$ is always integrable and simple to solve.  $A$ and $B$ both conserve linear and angular momentum, which we verified numerically.

We construct an integrator using 
\begin{subequations} 
\label{eq:integ}
\begin{align}
\label{eq:BAB}
	\Exp{h \hat{\tilde{H}}} &= \Exp{\frac{h}{2} \hat{B}} \Exp{h \hat{A}}{}\Exp{\frac{h}{2} \hat{B}}{}
	\qquad\text{or}\qquad \\
	\Exp{h \hat{\tilde{H}}} &= \Exp{\frac{h}{2} \hat{A}}{} \Exp{h \hat{B}}{}\Exp{\frac{h}{2} \hat{A}}{},
	\end{align}
\end{subequations}
where $h$ is a timestep.  \cite{C99} and \cite{Reinetal2019} used the first version, $BAB$.  We call the second version $ABA$.  The $n = 1$ planet problem (the Kepler problem) is not solved exactly by these integrators.  To see this, let the maximum $i$ in Eqs. \eqref{eq:Hsplit} and \eqref{eq:defterms} be $1$.  This problem has a solution, involving modifying the terms placed in $A$ and $B$.  {\cite[e.g.][Section 4.2]{HD17}} showed how to do this when all $K = 0$, but the generalization to allow other $K$ is immediate.  We tested this alternative splitting, but it did not perform better in the tests in this paper because solving the $n = 1$ problem is not the only consideration for a good integrator \citep[Figure 6, and error analysis]{HD17}. $\tilde{H}$ is given by the BCH formula {\citep[e.g.][]{hair06}} as,
\begin{eqnarray}
	\label{eq:WH:Herr:BAB}
	\tilde{H}^{\mathrm{BAB}} &=&
	A + B 
	+\frac{h^2}{12}\{\{B,A\},A\}
	-\frac{h^2}{24}\{\{A,B\},B\}
	+\mathcal{O}(h^4),
	\\
	\label{eq:WH:Herr:ABA}
	\tilde{H}^{\mathrm{ABA}} &=&
	A + B
	-\frac{h^2}{24}\{\{B,A\},A\}
	+\frac{h^2}{12}\{\{A,B\},B\}
	+\mathcal{O}(h^4).
\end{eqnarray}
$\{\}$ are Poisson brackets {\citep[][Section 9.5]{gol02}}.  If all $K$ are 1, the Poisson bracket terms are given by \cite{HD17}:
\begin{subequations} \label{eq:err:WHI}
\begin{align}
	\{\{A,B\},B\}
		&= \{\{V_\odot,T_0\},T_0\} + \{\{T_1,\Vp\},\Vp\} &\propto& \epsilon^3
	\qquad\text{and}\\
	\{\{B,A\},A\}
		&= \{\{T_0,V_\odot\},V_\odot\} + \{\{\Vp,T_1\},T_1\} && \nonumber\\
		&-	\{\{T_1,\Vp\},V_\odot\} - \{\{V_\odot,T_0\},T_1\}  &\propto& \epsilon^2.
\end{align}
\end{subequations}
{$\epsilon = m/m_0$, where the largest planetary mass $m$ gives upper error bounds.  $\epsilon \ll 1$.}  If all $K = 0$, the Poisson brackets are,
\begin{subequations} \label{eq:err:K0}
\begin{align}
	\{\{A,B\},B\}
		&= \{\{V_\odot,T_0\},T_0\}  &\propto& \epsilon^3
	\qquad\text{and}\\
	\{\{B,A\},A\}
		&= \{\{T_0,V_\odot\},V_\odot\}  - \{\{V_\odot,T_0\},T_1\}  &\propto& \epsilon^2.
\end{align}
\end{subequations}
  Thus, the $ABA$ form has a lower error in the two limits.  We performed numerical experiments of this prediction by integrating a system composed of the Sun, Jupiter, and Saturn.  We tested three cases: $K$ was always 0, $K$ was always $1$, and $K$ allowed to enter $0 < K < 1$, using a $C^2$ (see Section \ref{sec:numex}) $K$.  We used a variety of stepsizes and integration times.  In all cases, the absolute magnitude of the energy error was about twice smaller in the $ABA$ case, as compared to $BAB$.  {This is consistent with \eqref{eq:WH:Herr:BAB} and \eqref{eq:WH:Herr:ABA}, where the magnitude of the $\{\{B,A\},A\}$ coefficient is twice smaller for $ABA$.}  $ABA$ is approximately just as expensive as $BAB$ because evolving the final $A$ from a step can be combined with evolving the first $A$ of the next step.  In this paper, we use the $BAB$ integrator form in our tests, but we note that the efficiency can be improved by using $ABA$.  The goal of this paper is not to optimize efficiency.  The equations of motion for Hamiltonian $A$ are \citep{Reinetal2019},

\begin{eqnarray}
    \v{  \dot Q}_i &=& \v V_i\\
    \v{  \dot V}_i &=& - \frac{Gm_0}{Q_i^3} \v Q_i \nonumber\\
             && - G \sum_{j\neq i, j>0} \frac{ m_j}{Q_{ij}^3} \v Q_{ij} \left(1-K(Q_{ij}) + Q_{ij}  K'(Q_{ij})\right), \label{eq:eomhkh}
\end{eqnarray}
where $\v{V}_i = \v{P}_i/m_i$ is the barycentric velocity.  Using velocities instead of momenta is one way to ensure we can describe test particles.  We used \citep{Reinetal2019},
\begin{eqnarray}
    \frac{\partial K(Q_{ij})}{\partial \v{Q}_{ij}} = \left.\frac{\v Q_{ij}}{Q_{ij}}\, \frac{\partial K(r)}{\partial r} \right|_{r=Q_{ij}}
    = \frac{\v Q_{ij}}{Q_{ij}}\, K'(Q_{ij}).
\end{eqnarray}
The equations for Hamiltonian $B$ are,
\begin{eqnarray}
\label{eq:test}
    \v{  \dot Q}_i &=& \frac{1}{m_0} \sum_{i \ne 0} \v{P}_i \\
    \label{eq:test1}
    \v{\dot V}_i &=&  - G \sum_{j\neq i, j>0} \frac{m_j}{Q_{ij}^3} \v Q_{ij} \left( K(Q_{ij}) 
             -  Q_{ij} K'(Q_{ij}) \right) \label{eq:eomhik}
\end{eqnarray}
Since $\ddot{\v{Q}}_i$ is $0$, $\dot{\v{Q}}_i$ is a constant.  $\dot{\v{V}}_i$ is constant because $\dot{\v{Q}}_{ij}$ is $0$.  

\section{Modified differential equations}
\label{sec:MDE}
Although we have written the equations of motion for $A$ and $B$, the integrators of \eqref{eq:integ} can be described by sets of ODEs called modified differential equations (MDEs) {\citep{hair06,HB18}}.  We derive these for the first-order integrator,
\begin{eqnarray}
\label{eq:AB}
    \Exp{h \hat{\tilde{H}}} = \Exp{{h} \hat{A}} \circ \Exp{h \hat{B}}.
\end{eqnarray}
The Section \ref{sec:example} example also uses a first order integrator.  Here,
\begin{equation}
\tilde{H} = T_1 + T_0 + V_{\sun} + \sub{V}{pl} +  \sub{V}{p2}+ \frac{h}{2} \left( \{\sub{V}{p2},T_1 \} + \{T_0, V_{\sun} \} \right) + \mathcal O (h^2).
\end{equation}
The set of ODEs comes from $\tilde{H}$.  For $i > 0$,
\begin{subequations}
\label{eq:test}
\begin{align}
    \v{  \dot Q}_i &= \frac{\partial \tilde{H}}{\partial \v{P}_i} = \v{a}_{0i} + \v{a}_{1i} h + \mathcal O(h^2), \qquad\text{and}\\
    \v{\dot V}_i &= - \frac{1}{m_i}\frac{\partial \tilde{H}}{\partial \v{Q}_i} = \v{b}_{0i} + \v{b}_{1i} h + \mathcal O(h^2),
              \end{align}
              \label{eq:eomhik}
\end{subequations}
We defined,
\begin{subequations}
	\label{eqs:e}
\begin{align}
	\v{a}_{0i} &= \frac{\v{P}_i}{m_i} + \frac{\v{P}_{\sun}}{m_0}, \\
	\v{a}_{1i} &= \frac{1}{2} \left( \sum_{j \ne i, j>0} \frac{G m_j}{Q_{i j}^3} \v{Q}_{i j} (K(Q_{ij}) - Q_{i j} K^\prime(Q_{ij})) +  \sum_{j > 0} \frac{G m_j}{Q_j^3} \v{Q}_j \right), \\
	\v{b}_{0i} &= - \frac{G m_0}{Q_i^3} \v{Q}_i - \sum_{j \ne i, j> 0} \frac{G m_j}{Q_{i j}^3} \v{Q}_{i j}, \qquad \text{and} \\
	\v{b}_{1i} &= -\frac{1}{2} \left( \sum_{j \ne i, j > 0} \frac{G m_j}{Q_{i j}^5} \left[ (-3 \v{Q}_{i j} (\v{Q}_{i j} \cdot \v{V}_{i j})) (K(Q_{ij}) - Q_{ij} K^\prime(Q_{ij}) \right. \right. \nonumber \\
	& \left. \left.+ 1/3 Q_{ij}^2 K^{\prime\prime}(Q_{ij})) - \v{V}_{i j} Q_{i j}^2 (K(Q_{ij}) - Q_{i j} K^\prime(Q_{ij}))   \right] \right. \nonumber \\
	 & \left.+ \frac{G}{Q_i^5} \left[ -3\v{Q}_i  (\v{Q}_i \cdot \v{P}_{\sun} )+ \v{P}_{\sun} Q_i^2  \right]\right), \nonumber \\
\end{align}
\end{subequations}
where $\v{P}_{\sun} = \sum_{i > 0} \v{P}_i$ \citep{HD17}. Eqs. \eqref{eq:eomhik} are not Taylor series in $h$: $h$ is a constant parameter.  The MDEs do not need to converge in $h$, they correctly describe the dynamics.  Eqs. \eqref{eqs:e} are generally nonzero for massive and massless particles.  $\v{a}_{0 i}$ is the heliocentric velocity and $\v{b}_{0i}$ is the Newtonian force: in the limit $h \to 0$, these are the only surviving terms and the MDEs exactly solve the gravitational $N$-body equations.  We can see that the integrator \eqref{eq:AB} suffers from the same problem as integrators \eqref{eq:integ}, and any integrator split into the $A$ and $B$ from \eqref{eqs:split1}: the $n = 1$ problem is not solved exactly, even when $K(Q_{ij})$ and its derivatives are 0.  We see this because  terms $\v{a}_{1 i}$ and $\v{b}_{1i}$ are nonzero, although $\v{a}_{1 i} = 0$ for test particles.  The coefficients as the power of $h$ increases depend on higher derivatives of $K$.  Up to first order in $h$, two derivatives of $K$ are needed.  As we integrate these MDEs over discontinuities in those derivatives, unresolved timescales can occur \citep{H19}, leading to large errors.  As $h$ is increased, these discontinuities will become more important.  The MDEs corresponding to integrator $BAB$, Eq. \eqref{eq:BAB}, also the one considered in \cite{C99} and \cite{Reinetal2019}, do not have form \eqref{eq:test}; the leading error term is $\mathcal O (h^2)$.  However, Eqs. \eqref{eq:test} and \eqref{eqs:e} let us see how discontinuities would enter any MDEs.  
\subsection{Example error analysis}
\label{sec:example}
\begin{figure}
	\includegraphics[width=80mm]{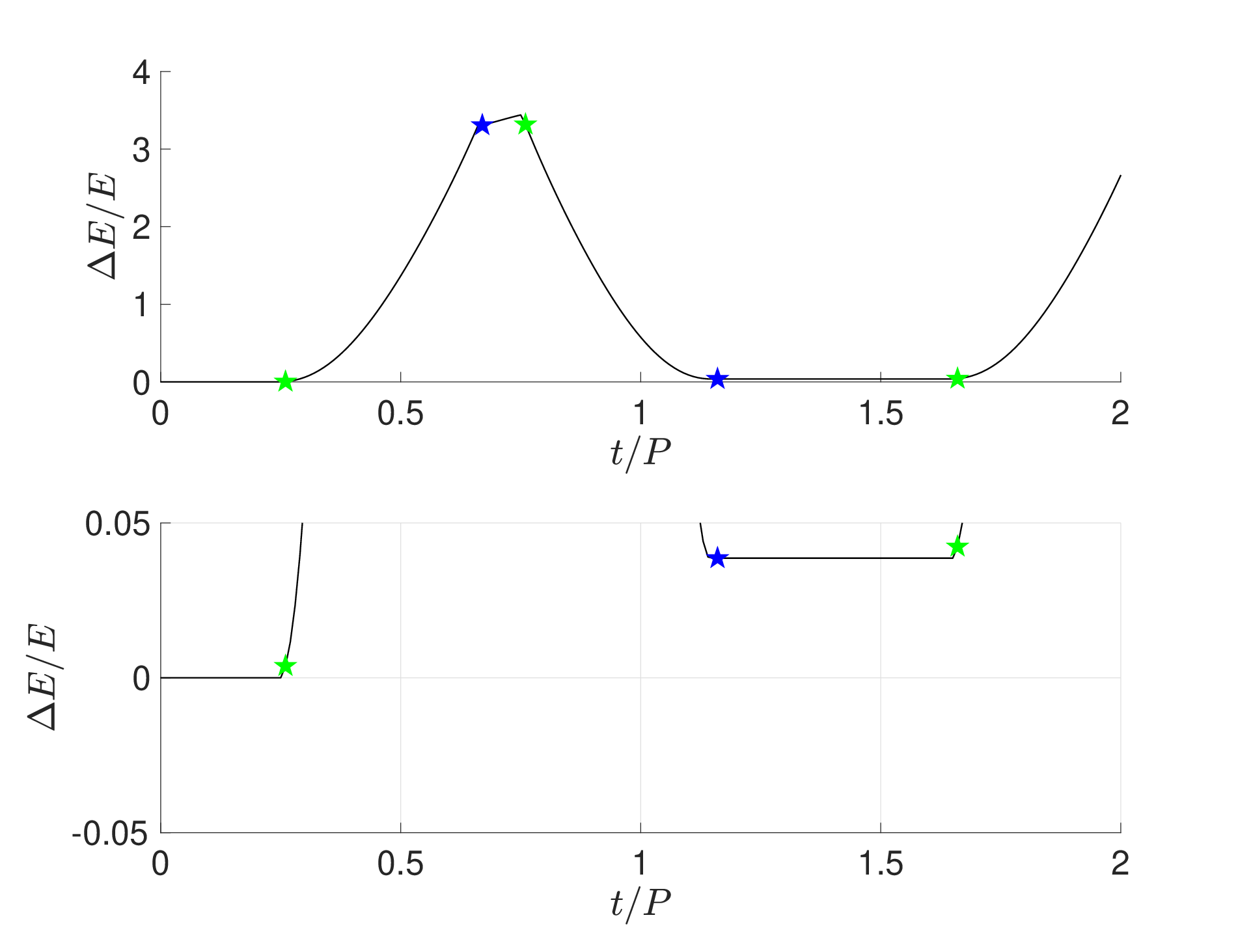}
	\caption{Energy error as a function of time for an integration with the SHO integrator, \eqref{eq:shoint}.  Initially, $q = 1$ and $p = 0$.  A time step $h = P/100$ is used, where $P = 2 \pi$ is the exact SHO period.  Four steps integrate over a discontinuity {in $dK/dr$} in the first period--- they are indicated with stars.  A green star locates the end of a time step where an integration out of the region $0 < K < 1$ was performed.  A blue star indicates an integration into the region $0 < K < 1$.  The bottom panel is a zoomed in region of the top panel.  In it, the floor of the energy error shifts $\approx +0.035$.  This is explained with error analysis in Section \ref{sec:example}.
	\label{fig:shodisc}
  	}
\end{figure}

Here we show and explain an example of {a finite DCO $K$} giving rise to secular errors {in orbits}.  This subsection can be omitted without loss of continuity.  Consider the simple harmonic oscillator (SHO), with Hamiltonian,
\begin{equation}
H = \frac{p^2 + q^2}{2}.
\end{equation}
We are concerned in this work with chaotic orbits, but for this example, we {consider} a periodic orbit. We construct a hybrid integrator to solve this problem.  This integrator was considered in \cite[Section 4.1]{H19}.  The Hamiltonian is split into,
\begin{equation}
H_1 = \frac{q^2 K(q) + p^2}{2} \quad \text{and} \quad H_2 =   \frac{{q^2} (1 - K(q))}{2}.
\end{equation}
The integrator is,
\begin{equation}
\label{eq:shoint}
\Exp{h \hat{\tilde{H}}} = \Exp{h \hat{H_2}} \circ \Exp{h \hat{H_1}}.
\end{equation}
$H_1$ is solved using Bulirsch--Stoer (as in \cite{H19}).  Let $x = (q+2)/2$.  If $x < 0$, $K(q) = 0$.  If $x > 0$, $K(q) = 1$.  Otherwise, $K(q) = x$.  This $K$ is a $C^0$ function, so integrator \eqref{eq:shoint} is not symplectic, according to \cite{H19}.  As in \cite{H19}, initially, $q = 1$ and $p = 0$. $h = P/100$, where the exact period is $P = 2 \pi$.  The runtime is $t = 2 P$.  The energy error as a function of time is plotted in Fig. \ref{fig:shodisc}.  

In the first period, four steps, shown with stars, integrate over a discontinuity in $dK/dr$.  Stars locate the end of those steps.  A green star means the region $0 < K < 1$ was left, while a blue star means $0 < K < 1$ was entered.  The bottom subpanel zooms in on a portion of the top subpanel, and shows the energy error floor shifting.  We now explain the magnitude and sign of this shift.  A succession of such shifts lead to long-term error deterioration.  

First, we calculate $\tilde{H}$, given by the BCH formula.  If the series converges, we can ignore high powers in $h$ and assume $\tilde{H}$ is conserved.  We have,
\begin{equation}
\label{eq:htild}
\tilde{H} = H + \frac{h}{2} \left\{H_1, H_2 \right\} + \mathcal O(h^2),
\end{equation}
The Poisson brackets are calculated as,
\begin{equation}
\label{eq:pbrack}
\{H_1,H_2\} = p \left( \frac{q^2}{2} \frac{dK}{dq} - q (1 - K) \right).
\end{equation}

Note that $\tilde{H}$ describes exact, continuous dynamics, explained by MDEs, although the integrator samples discrete points along the solution path.  We calculate the error across a discontinuity.  Using \eqref{eq:htild} and \eqref{eq:pbrack}, and ignoring higher $h$ powers,
\begin{equation}
\delta = \frac{\Delta H}{H_0} = \frac{-\frac{h}{2} \left(\{H_1,H_2\}\bigg\rvert_{+} - \{H_1,H_2\} \bigg\rvert_{-}\right)}{H_0} = s a p q^2.
\end{equation}
$H_0$ is the initial energy $(1/2)$.  $\bigg\rvert_{+}$ indicates the limit from the right (forward time) of the discontinuous point while $\bigg\rvert_{-}$ is the limit from the left.  $a= h/(8 H_0)$.  $s$ is $+1$ if $0 < K < 1$ was entered while it is $-1$ if $0 < K < 1$ was left.  We calculate the first four $\delta$ values of Fig. \ref{fig:shodisc} approximately, because only data at discrete values is available from the integrator.  We use the coordinates after integrating over the discontinuity.  Labelling each $\delta$, $\delta_1 = -6.2 \times 10^{-5}$, $\delta_2 = 0.027$, $\delta_3 = 0.039$, and $\delta_4 = -5.8 \times 10^{-5} $.  The maximum $\delta$ and maximum $p q^2$ are, respectively, $0.039$ and $\approx 2.5$.  The jump in the bottom subpanel of Fig. \ref{fig:shodisc} is $+ 0.035$, so this error analysis provides an approximation for this jump.  Reducing the step by 10 to $h = P/1000$, there are again four discontinuities over the first period.  The maximum magnitude of $p q^2$ is similar, $\approx 2.3$, and $a$ is $10$ times smaller.  The error floor jumps in this case by $ + 0.004$, {so} the error analysis result again matches approximately the data.

\section{Numerical experiments}
\label{sec:numex}
\subsection{Chaotic orbits}

\begin{figure}
	\includegraphics[width=80mm]{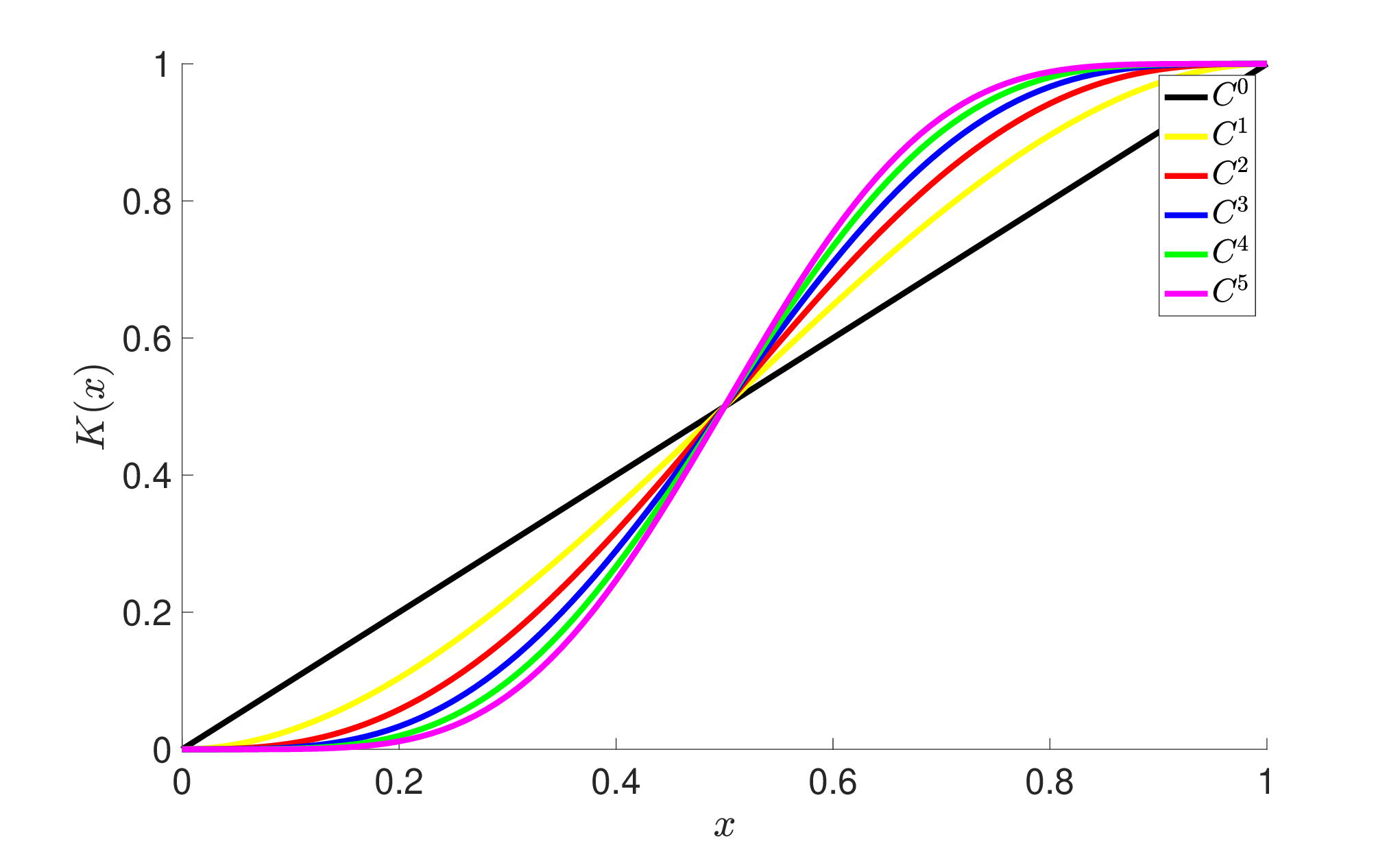}
	\caption{We plot the transition functions considered in this work.  The functions are labelled $C^n$, where $n$ is the number of derivatives that are continuous.  The functions are all {$C^\infty$}, except at $x = 0$ and $x = 1$.  The linear $C^0$ function is continuous, but not differentiable. 
	\label{fig:Kx}
  	}
\end{figure}

We return attention to the $N$-body problem. We have yet to specify the functional form of $K$.  To discuss $K$, first consider a function $x = (r - h_1 R)/(h_2 R)$, {a} function {used} by \cite{Wisdom2017}.  $r$ is the separation between two non-{stellar} bodies, and $R$ is a constant.  If $x < 0$, $K = 0$.  If $x > 1$, $K = 1$.  Otherwise, we {use} the following progressively smoother functions,
\begin{enumerate}
\item{$C^0$: $K = x$}
\item{$C^1$: $K = -2x^3 + 3x^2$}
\label{it:c1}
\item{$C^2$: $K = 6x^5 - 15x^4 + 10x^3$}
\label{it:c2}
\item{$C^3$: $K = -20x^7 + 70x^6 - 84x^5 + 35x^4$.}
\item{$C^4$: $K = 70x^9 - 315x^8 + 540x^7 - 420x^6 + 126x^5$}
\item{$C^5$: $K = -252x^{11} + 1386x^{10} - 3080x^9 + 3465x^8 - 1980x^7 + 462x^6$}
\end{enumerate}
The functions increase monotonically and have odd-symmetry about $x = 1/2$.  They are displayed in Fig. \ref{fig:Kx}.
{\cite{Wisdom2017} used a $C^2$ function and \cite{C99} used a $C^1$ function.  The code \texttt{MERCURY}, however, described in \cite{C99}, uses a $C^2$ function.  Thus, the forces in \texttt{MERCURY} are smoother than those of \cite{Wisdom2017}, although the former has symplecticity and time-reversibility errors \citep{H16}.}

Following \cite{Wisdom2017}, if, for a given pair {of non-stellar objects}, $x > h_3 R$ at the start of $A$, we use Kepler solvers to solve the Hamiltonian terms with their indices $i$ and $j$.  $h_3 R > 1$ by construction.   Otherwise, Bulirsch--Stoer is used for that pair in $A$, to account situations in which $K$ becomes less than $1$ during the evolution of the $A$ Hamiltonian.

We consider a simple chaotic system with two degrees of freedom, the restricted three-body problem.  It has one constant of motion.  In an inertial frame it is,
\begin{equation}
\label{eq:jac}
	C_J
	= \tfrac{1}{2}\vec{v}^2 + \Phi(\vec{x}) - \vec{\omega}\cdot(\vec{x}\times\vec{v}),
\end{equation}
where $\omega$ is the binary angular velocity, $\vec{\omega}=\sqrt{G(m_1+m_2)/a^3}\hat{\vec{z}}$, and
\begin{equation}
	\Phi(\vec{x}) = -\frac{Gm_1}{|\vec{x}-\vec{x}_1|}
					-\frac{Gm_2}{|\vec{x}-\vec{x}_2|}.
\end{equation}
$a$ is the semi-major axis {of the circular orbit}, $m_1$ and $m_2$ are {the masses}, and $\v{x}_1$ and $\v{x}_2$ are their positions.  {$m_1$ is {the} star}.  $\v{x}$ and $\v{v}$ are the position and velocity of the test particle.  In units of au, day, and solar masses, the test particle has heliocentric position and velocity 
\begin{equation}
\label{eq:IC}
\begin{aligned}
\v{x} &= (4.42,0) \\
\v{v} &= (0, 0.0072).
\end{aligned}
\end{equation}
$m_1 = 1$ and $m_2$ is defined through $\mu = m_2/(m_1+m_2) = 0.01$.  $a = 5.2$.  Initially, all bodies are aligned with the $x$-axis, with the star on the left.  The center of mass is at $0$.  This problem was {analyzed} by \cite{Wisdom2017}, so a comparison is possible in integrator performance.  We let $R$ be the Hill radius, $R = a (m_2 /(3 m_1))^{1/3}$, $h_1 = h_2 = 1.5$, and $h_3 = 4.0$, as in \cite{Wisdom2017}.  For all tests in this paper, we checked that when Kepler solvers were used to solve $A$, a close encounter condition, $K < 1$, was not detected after evolving the $A$ Hamiltonian.  This justifies the choice of $h_3$.

\begin{figure}
	\includegraphics[width=80mm]{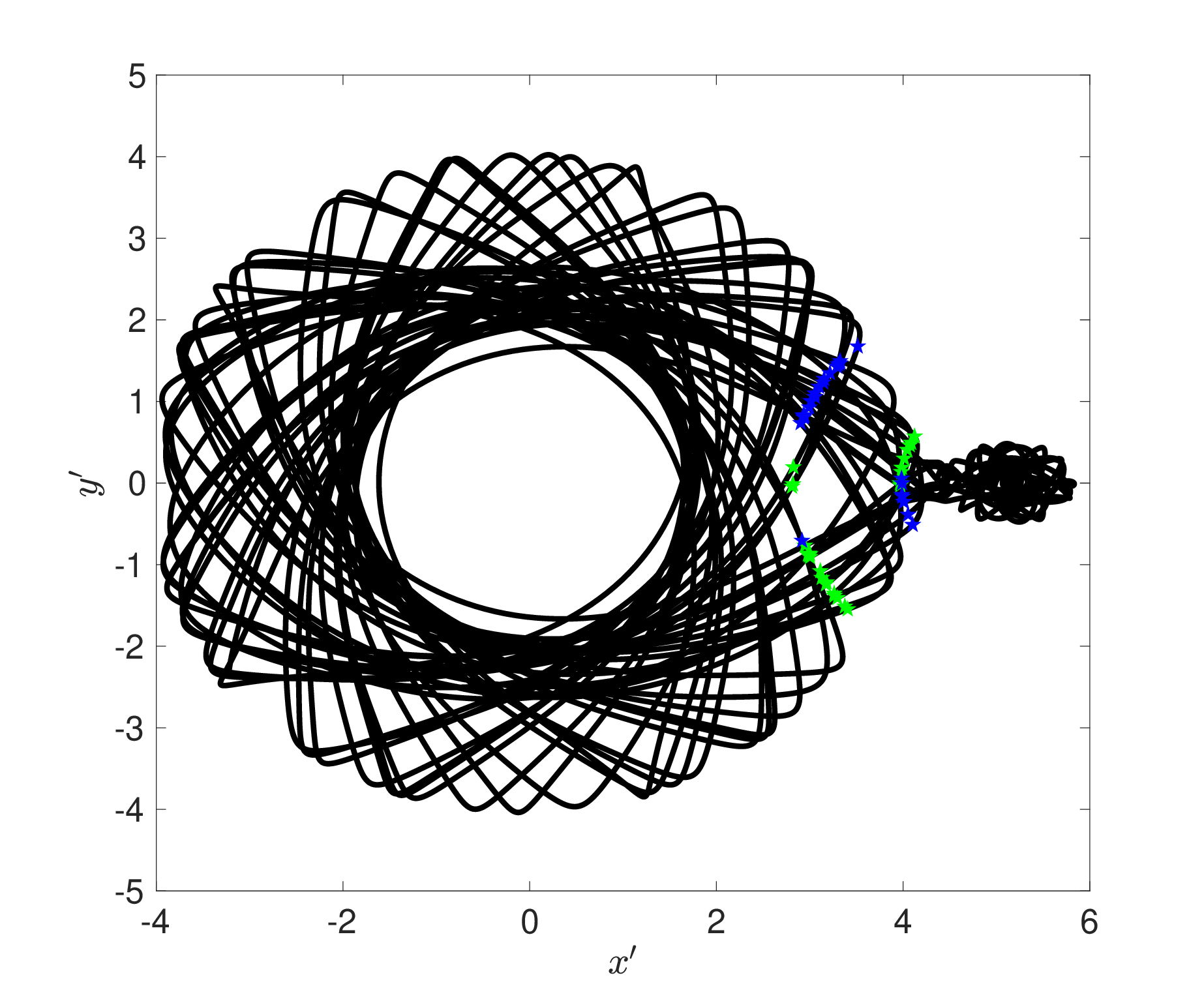}
	\caption{The trajectory in the rotating frame of the test particle in the restricted three-body problem described in Section \ref{sec:numex}.  The initial conditions are \eqref{eq:IC}.  The $BAB$ integrator is used with $h = 8$ days and the $C^2$ transition function.  The total runtime is $500$ yrs.  Green and blue stars indicate a step with an integration over a discontinuity.  Green stars mean the test particle left the region $0 < K < 1$, while blue stars mean the particle entered the $0 < K < 1$ region.  Note we have not used red to draw the $C^2$ curve as in Fig's. \ref{fig:Kx}, \ref{fig:smooth}, \ref{fig:changeic}, and \ref{fig:compare}. 
	\label{fig:traj}
  	}
\end{figure}

Using {function \ref{it:c2}}, $h = 8$ days, and running for a total time $t = 500$ yrs, we show the trajectory in the rotating frame in Fig. \ref{fig:traj}.
We have also indicated with green stars if a transition out of the region $0 < K < 1$ happened during the timestep yielding the coordinates.  This means an integration over a discontinuity was performed.  We also show with blue stars when a transition into $0 < K< 1$ happened.  $54$ steps integrated over a discontinuity, out of a total of $22,830$ steps.  Transitions should occur at separations of $r = 3.0 R$ and $r = 1.5R$, respectively, which we verified numerically.

\begin{figure*}
	\includegraphics[width=120mm]{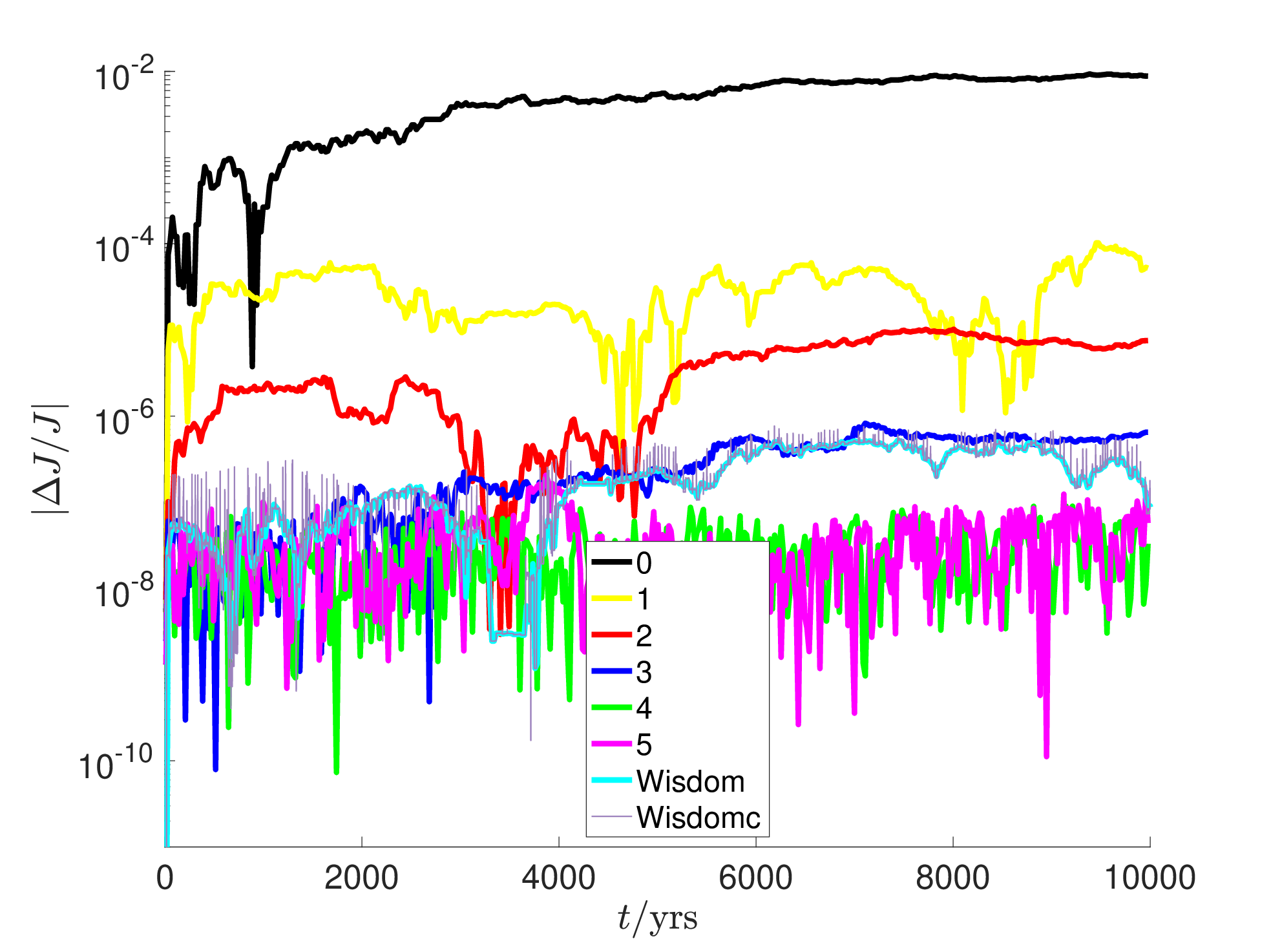}
	\caption{The absolute value of the error of the Jacobi constant, eq. \eqref{eq:jac}, for the restricted three-body problem considered by \protect\cite{Wisdom2017}.  {First, t}he $BAB$ integrator with step $h = 8$ days is used with different transition functions.  The legend denotes {the differentiability class order of the smoothing function}.  We have also plotted data from \protect\cite{Wisdom2017}, Figure{s} 9 {and 10}, {respectively, as ``Wisdom'' and ``Wisdomc.''  The data were }kindly provided by Jack Wisdom.  {Except for the last curve, all data has been processed, as described in the text, to reduce error oscillations.}  Increasing {smoothness} improves the error by nearly five orders of magnitude and {also} improves on the \protect\cite{Wisdom2017} data.
	\label{fig:smooth}
  	}
\end{figure*}

Next, we measure the Jacobi constant error for long-term integrations over $10,000$ yrs using various smoothing functions.  We expect that smoother functions will have better error.  We use $h = 8$ days, but to reduce variations in the error, we plot the median error every 1000 steps, so that a total of 456 points are plotted.  {The result is in Fig. \ref{fig:smooth}.}  We also plot data from {Figure 9 of} \cite{Wisdom2017} {(``Wisdom'')}.  For these data, median absolute errors every $20$ steps are plotted, so that there are also a total of $456$ points in the range $t = 0$ to $t = 10,000$ yrs.  {We also plot corrected data from Fig. 10 of \cite{Wisdom2017} as ``Wisdomc.''  These data behave similarly to the median errors from above, and the two curves lie on top of each other.}  We indicate by integers $n$ the {DCO.}   

Integrations with smoother transition functions have smaller error, although there is no significant improvement from $n = 4$ to $n = 5$.  We can improve on the \cite{Wisdom2017} data simply by increasing smoothness at virtually no additional cost.  Running to 500 yrs, the $C^5$ experiment took about $1\%$ longer in wall-clock time than the $C^2$ test.  The $C^5$ test was faster than the $C^0$ test, but note different runs may have a different number of evaluations of the polynomial function.  We used the MATLAB software for this timing test.  But our goal is not to improve on the \cite{Wisdom2017} result--- if our focus were on accuracy, we could reduce the error on all our curves by using instead the $ABA$ forms.  Our goal is to demonstrate the importance of the transition function.  

\begin{figure*}
	\includegraphics[width=120mm]{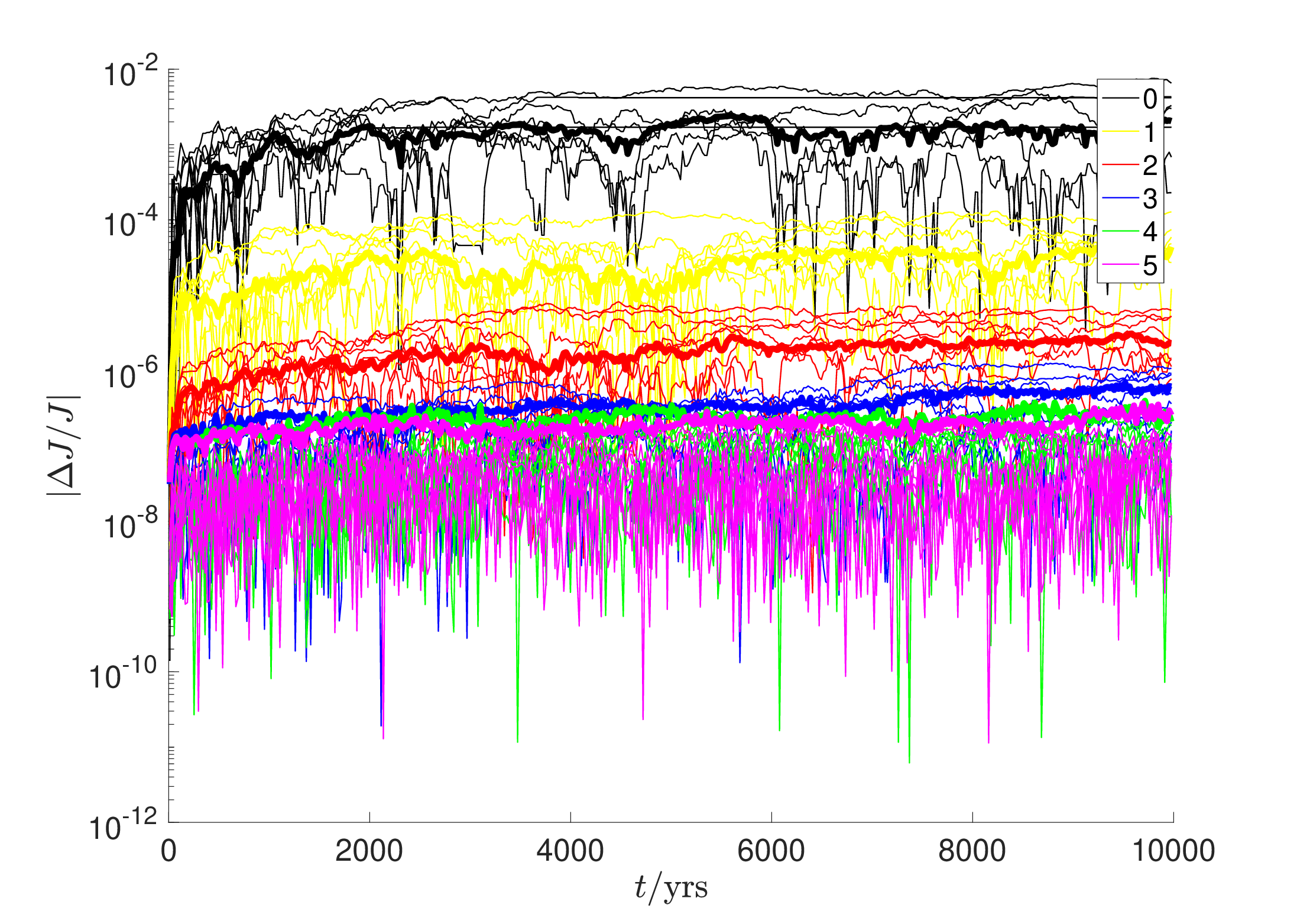}
	\caption{The same test is run as in Fig. \ref{fig:smooth}, but for each transition function, $8$ initial conditions are run.  Each is indicated with a thin line.  Thick lines indicate the geometric mean of the thin lines.  The improvement in error with smoothness observed in Fig. \ref{fig:smooth} is a robust result that does not depend on a particular choice of initial conditions.
	\label{fig:changeic}
  	}
\end{figure*}

Because the error as a function of time strongly depends on the initial conditions, we want to see if our results hold for a variety of orbits.  Thus, we perturb the initial conditions \eqref{eq:IC}, but not so much that the initial conditions' Jacobi constant no longer satisfies the condition for a chaotic exchange oribt \citep[Section 3.3]{DH17}.  For the test of Fig. \ref{fig:traj}, the variation in either test particle position coordinate is $\approx 10$, while the variation in either velocity is $\approx 0.03$.  Thus, we set a position and velocity scale as $x_{\text{s}} = 10$ and $v_{\text{s}} = 0.01$, respectively.  Then $8$ initial conditions are generated by perturbing one of the positions or velocities at a time from the initial conditions above.  The positions are perturbed by $\pm  x_{\text{s}} \delta$ and velocities by $\pm v_{\text{s}} \delta$, where $\delta  = 10^{-3}$.  For the long timescales we study in our tests, we can perturb initial conditions even by small amounts like $10^{-15}$.  Assuming a Lyapunov time of $4$ yrs, if we perturb the initial conditions by $10^{-15}$, the trajectories will differ by unity after $\approx 136$ yrs.  However, in one such test, significant differences between orbits were not apparent until after $400$ yrs.  We repeat the test from Fig \ref{fig:smooth}, but with $8$ initial conditions with each transition function.  We plot their errors by again taking median values and using thin lines.  A thick line represents an average in log space, or the geometric mean, of the eight curves.  The result is shown in Fig. \ref{fig:changeic}.
This plot confirms the improvement in error from increasing smoothness.  As in Fig. \ref{fig:smooth}, there is improvement in error until $n = 4$.  It is unclear whether this is an improvement from $n = 4$ to $n = 5$.

We verified the orbits from nearby initial conditions diverge quickly.  For a certain integrator and the unperturbed initial conditions \eqref{eq:IC}, \cite{DH17} reported a Lyapunov time of $4$ yrs.  For the $n = 2$ case, we compared the orbit in which the $x$ coordinate is perturbed by $- x_{\text{s}} \delta$ and the orbit for which $y$ is perturbed by $+x_{\text{s}} \delta$.  We calculated the Euclidean norm of the distance between the two orbits, which reached unity before $5$ yrs.  The Euclidean norm in velocity space reached its saturation of approximately $10^{-2}$ at a similar time.

\begin{figure*}
	\includegraphics[width=120mm]{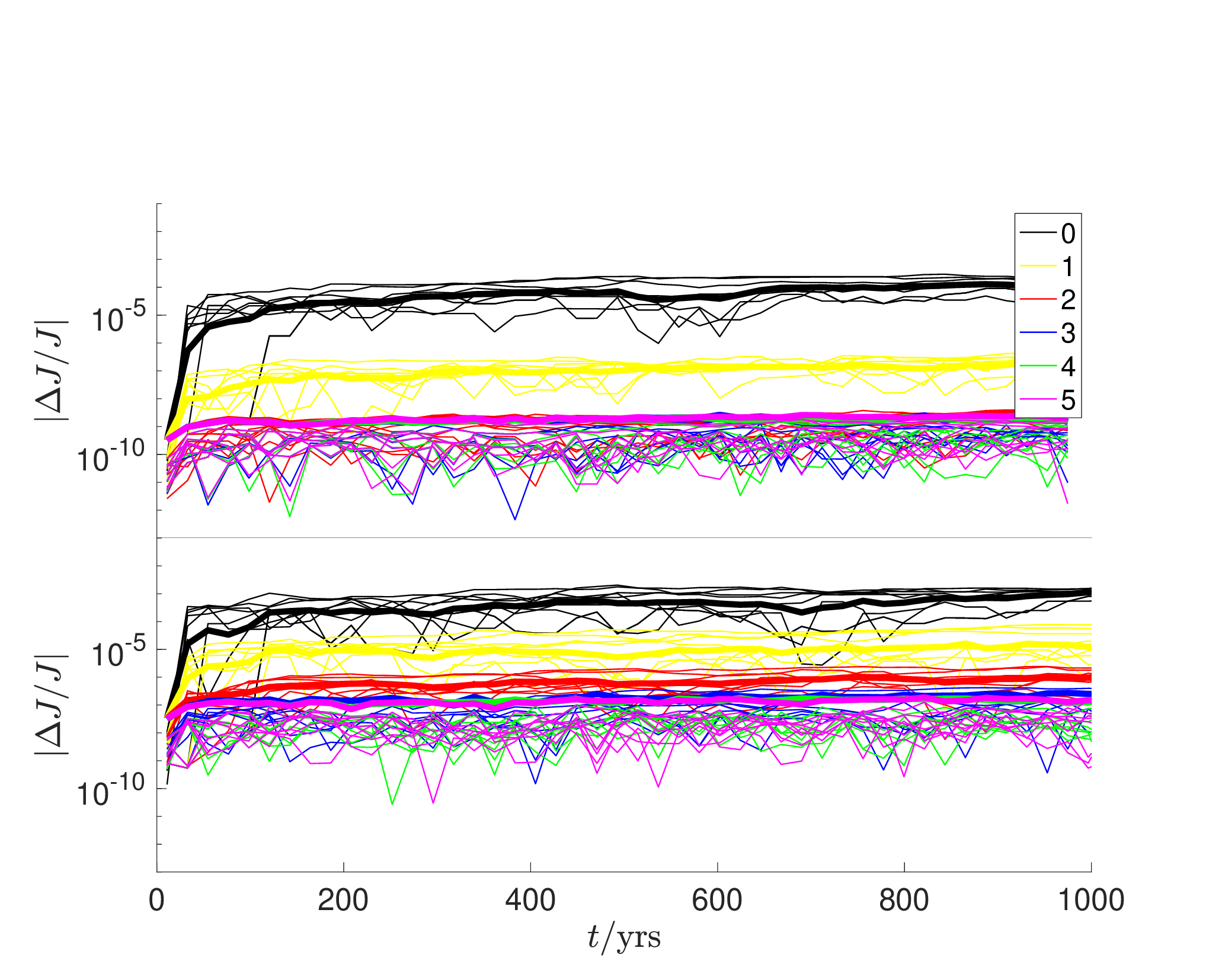}
	\caption{We compare how the impact of smoothness on error depends on the time step.  The bottom panel uses data from Fig. \ref{fig:changeic}.  The top panel is the same integration with step 10 times smaller, $h = 0.8$ days.  As predicted, smoothness makes less of an impact in the top panel.   Note the runtimes are only $1,000$ yrs.
	\label{fig:compare}
  	}
\end{figure*}

Based on the MDEs from \eqref{eq:test}, we expect that as $h$ is decreased, the smoothness of the transition will matter less, as its impact on the error becomes smaller.  To show this, we rerun the initial conditions \eqref{eq:IC} and their $8$ perturbed orbits using a step $10$ times smaller, $h = 0.8$ days.  Median values are taken every $10,000$ steps for plotting, so that the same number of points are plotted as in Fig. \ref{fig:changeic}.  We compare the Jacobi constant errors in Fig. \ref{fig:compare} between the two timestep runs, up to $t = 1,000$ yrs.  The top panel uses $h = 0.8$ days.  There is no clear difference once $n > 1$ in this case, while for $h = 8$ days case, differences are hard to see for $n > 2$.  Also note all thick curves have smaller error for the $h = 0.8$ days case.

\begin{figure}
	\includegraphics[width=80mm]{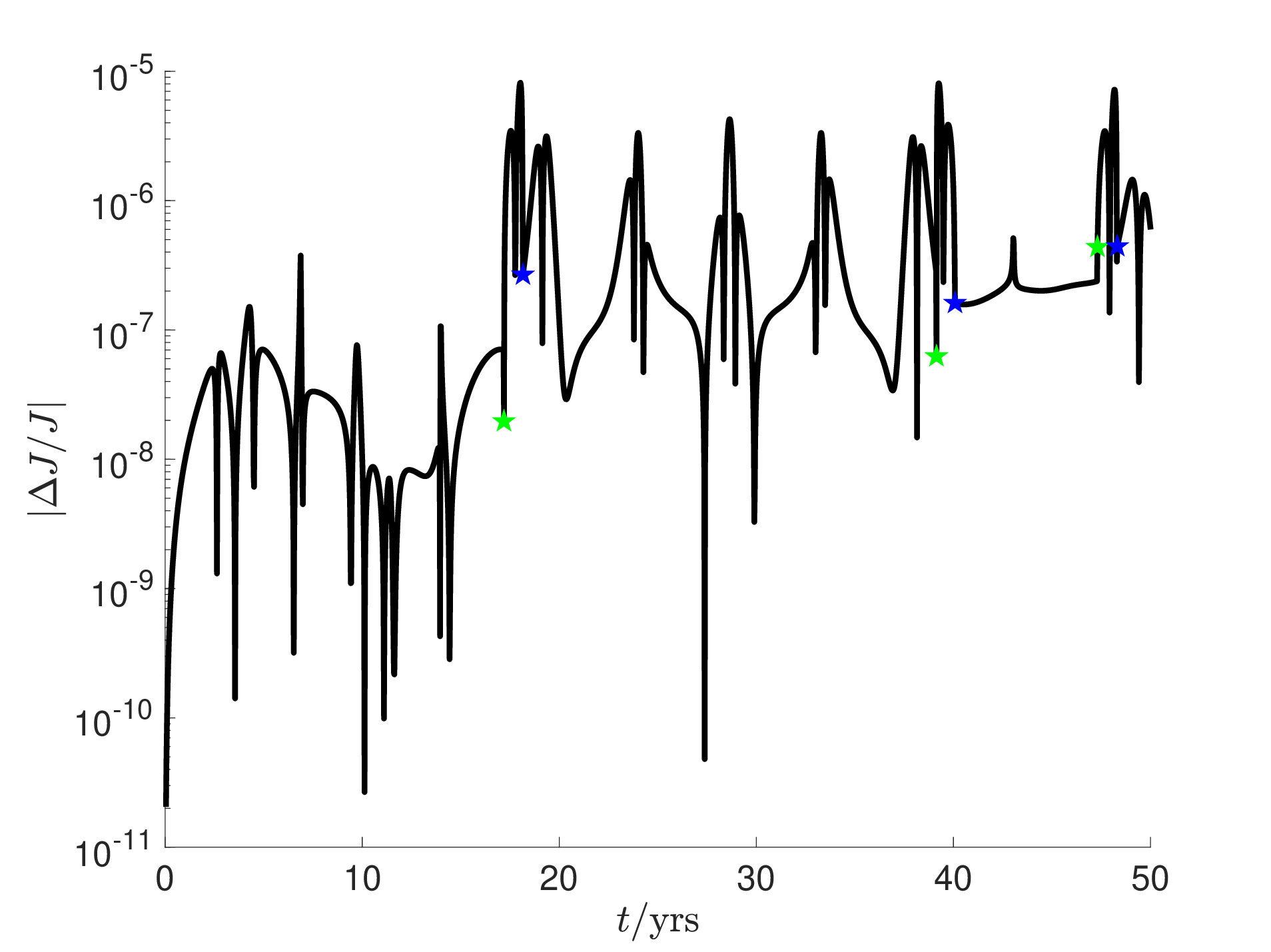}
	\caption{We show that integrations over discontinuities lead to error increases.  This plot takes the first $50$ years from the $C^2$ function in Fig. \ref{fig:smooth} {without computing medians.}  Stars indicate a step where an integration over a discontinuity occurred.  The meaning of the stars is the same as in Fig. \ref{fig:traj}.  Note we have not used red to draw the $C^2$ curve as in Fig's. \ref{fig:Kx}, \ref{fig:smooth}, \ref{fig:changeic}, and \ref{fig:compare}. 
	\label{fig:discont}
  	}
\end{figure}

{W}e want to demonstrate that large errors in the Jacobi constant are associated with integrations over discontinuities.  In Fig. \ref{fig:discont} we plot the error in the Jacobi constant for the $C^2$ function over the first $50$ yrs, using $h = 8$ days.  We also plot when an integration over a discontinuity occurred.  We find the first discontinuity is associated with a secular increase in the Jacobi constant error.  Further discontinuities are associated with other error spikes.

\subsection{{Revisiting regular orbits}}
\label{sec:conf}

{This work has focused on the effects of smoothness on chaotic orbits.  \cite{H19} showed that regular orbits were generally ensured with $C^{1,1}$ integrators.  We explore here whether increasing smoothness past $C^{1,1}$ improves the regular orbits.

We consider the Kepler orbit in \cite{H19} and use map (9) in that paper.  We run it for $t = 10000 P$, $h = P/100$, and eccentricity $e = 0.7$.  $P = 2 \pi$ is the period.  The initial conditions are at apoapse ($r = 1+e$, $p = 0$, using the same notation as \cite{H19}).  An error is calculated at each apocenter passage using a cubic polyynomial interpolation.  Specifically, assume apocenter passage occurs between $t_1$ and $t_1 + h$.  Then, apocenter occurs at,
\begin{equation}
r = a + b (t - t_1) + c \frac{(t-t_1)^2}{2} + d \frac{(t-t_1)^3}{6},
\end{equation}
where $a$, $b$, $c$, and $d$ are known constants and $t$ is found from $p = \dot{r} = 0$.  Fig. \ref{fig:confirm} plots the errors vs time for this experiment, where the smoothness of the map is varied.  The amplitudes for the $C^1$, $C^2$, and $C^\infty$ runs are, approximately, $1.9 \times 10^{-3}$, $9 \times 10^{-4}$, and $7 \times 10^{-4}$.  Thus, our chaotic results apply to regular orbits; error amplitudes decrease as smoothness is increased.  But a minimum smoothness is needed to guarantee the orbits are regular.
\begin{figure}
	\includegraphics[width=80mm]{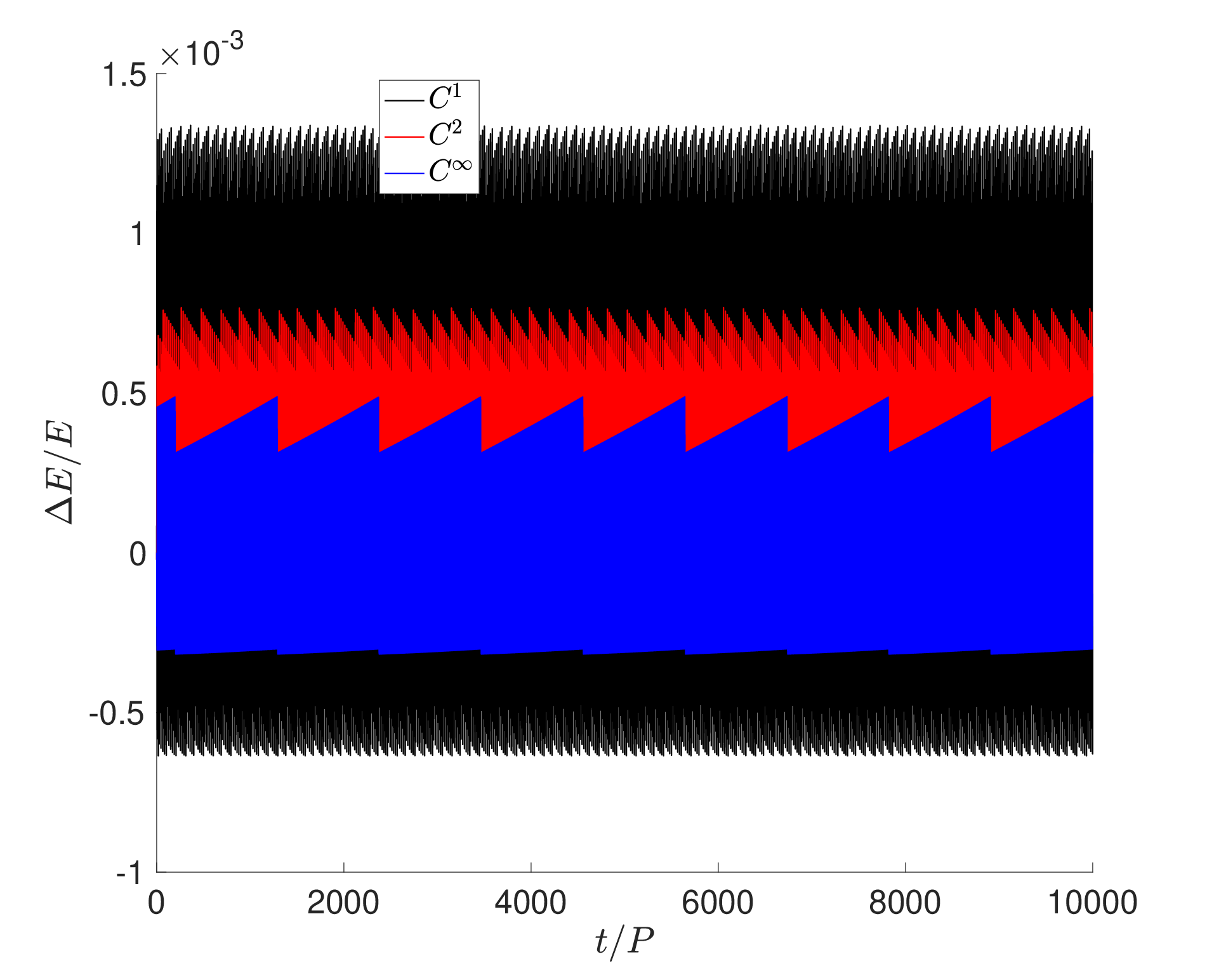}
	\caption{{Hybrid integrations of the Kepler problem using maps of different smoothness.  Errors are plotted at apoapse using polynomial extrapolation.  The error amplitude decreases as smoothness increases.}
	\label{fig:confirm}
  	}
\end{figure}

}

{We verified that using a given function as a force, rather than Hamiltonian, switching function, gives smaller errors.  This is expected as the force is obtained from differentiating the potential.  We also used other conventional integrators besides Bulirsch--Stoer which are less sensitive to smoothness, but found no significant impact to our results.}

\section{Conclusion}
\label{sec:conc}
Symplectic integration has been popular and successful at describing $N$-body {dynamics in astrophysics}.  Symplectic integration is part of a broader category of integration called geometric integration \citep{hair06} in which solutions to ordinary differential equations (ODEs) are calculated through approximations which respect {geometry} of the ODEs.  This is useful for any long-term dynamical studies.  Symplectic integration is supposed to conserve the Poincar\'{e} invariants associated with Hamiltonian dynamics, leading to bounded errors over long time intervals.  Popular symplectic integrators \mbox{\citep{WH91,S05}} also aim to preserve the other geometric features of the ODEs {such as} the time-symmetry, and the angular and linear momentum conservation.

Another geometric feature which is perhaps overlooked is the smoothness of an {ODE}.  The $N$-body Hamiltonian is infinitely smooth, or $C^\infty$ for physical separations greater than $0$.  Symplectic integrators are described by modified differential equations, which are also derived from a Hamiltonian with some smoothness.  A number of popular symplectic methods, and perhaps the most versatile and useful ones \mbox{\citep{C99,DLL98,S05}} are only $C^n$ smooth, where $n$ is a small integer, or worse, one might argue a smoothness concept is not applicable to them.  Any method with adaptive block stepping also has reduced smoothness.  This work has explored the role of smoothness in the accuracy of $N$-body integrations.  To explore this question, we have considered hybrid \citep{C99,Reinetal2019,Wisdom2017,H19} integrators, whose smoothness is easily tuned through a switching function.  Hybrid integrators are $C^\infty$ in most of phase space, except for a few points in phase space which are rarely or never hit, in which there is {a smaller differentiability class order (DCO)} (except if an expensive $C^\infty$ transition function is used).  However, \cite{H19} has shown that the performance of hybrid integrators degrades for regular orbits even when those points are never encountered in practice, and explained why.  

This work has focused on the broad class of chaotic orbits, and how their errors might be affected by smoothness of the method.  We have provided evidence that Hamiltonian smoothness has significant impacts on $N$-body performance.  In a test first considered by \cite{Wisdom2017} of a restricted three-body problem, the integrator performance significantly improved as $n$ was increased from $0$ to $4$ by about $5$ orders of magnitude in the Jacobi constant error.  We argued that this result can be explained through the MDEs.  We also verified the expectation that smoothness should become less important as the timestep goes to $0$.  {We also showed regular orbits were improved by smoothness.}  Our results indicate that just as using symplectic integrators has become a pillar of dynamical astronomy, using integrators that are as smooth as possible is a consideration which is arguably just as important.  Increasing smoothness{, or the DCO,} may not be easy to do in the case of block time-stepping methods, but for other methods, like hybrid integrators, increasing smoothness may be negligibly more computationally expensive, as we explored.

We remark on the relationship between smoothness and symplecticity:  \cite{H19} has argued that a symplectic integrator should have at least $C^{1,1}$ smoothness ($C^1$ and Lipschitz continuous in the ODEs), because these {appear to} guarantee the existence of periodic orbits.  This was verified numerically, and was predicted by \cite{AX16}, with some caveats described in \cite{H19}.  According to this criteria, we have explored both symplectic and non-symplectic integrators in this work, depending on the smoothness of the transition function.

Finally, we note that we have not investigated in detail the role smoothness plays in nonsymplectic integrators, but it is reasonable to assume smoothness can improve any integrator described by MDEs, symplectic or not.
\section{Acknowledgements}
I thank Jun Makino, Scott Tremaine, Matt Payne{, Jack Wisdom, and the anonymous referee}{s} for comments {that improved this work.}
\bibliographystyle{mnras}
\bibliography{paper}
\end{document}